# High-resolution investigation of spinal cord and spine

I. Bukreeva[1,2], V. Asadchikov[3], A. Buzmakov[3], V. Grigoryev[4], A. Bravin[5], A. Cedola[1]

[1] Institute of Nanotechnology- CNR, Rome Unit, Piazzale Aldo Moro 5, Italy

[2] P. N. Lebedev Physical Institute, RAS, Leninskii pr., 53 Moscow, Russia

[3] Shubnikov Institute of Crystallography FSRC "Crystallography and Photonics" RAS, Leninskii prosp., 59 Moscow, Russia

[4] National Research Nuclear University /Moscow Engineering Physics Institute, Kashirskoye Highway, 31 Moscow, Russia

[5] European Synchrotron Radiation Facility, 71 Avenue des Martyrs, 38043 Grenoble, Cedex France

## ABSTRACT

High-resolution non-invasive 3D study of intact spine and spinal cord morphology on the level of complex vascular and neuronal organization is a crucial issue for the development of treatments for the injuries and pathologies of central nervous system (CNS).

X-ray phase contrast tomography enables high quality 3D visualization in ex-vivo mouse model of both vascular and neuronal network of the soft spinal cord tissue at the scale from millimeters to hundreds of nanometers without any contrast agents and sectioning. Until now, 3D high resolution visualization of spinal cord mostly has been limited by imaging of organ extracted from vertebral column because high absorbing boney tissue drastically reduces the morphological details of soft tissue in image. However, the extremely destructive procedure of bones removal leads to sample deterioration and, therefore, to the lack of considerable part of information about the object.

In this work we present the data analysis procedure to get high resolution and high contrast 3D images of intact mice spinal cord surrounded by vertebras, preserving all richness of micro-details of the spinal cord inhabiting inside. Our results are the first step forward to the difficult way toward the high-resolution investigation of in-vivo model central nervous system.

# 1. INTRODUCTION

The spinal cord (SP) is a vital part of the human central nervous system (CNS), translating information between the brain and peripheral nervous system. The SC is protected by the vertebral column of the spine. The spine and the spinal cord are subject to a wide variety of diseases due to age, genetics, trauma, infection etc. Fracture of the spine and spinal cord injury (SCI) is the most devastating for the central nervous system and it causes the disability of patients mainly between the ages of 18 and 35 years. Worldwide, an estimated 2.5 million people live with spinal cord injury (SCI), with more than 130,000 new injuries reported each year [1]. The resulting injury to the central nervous tissue (CNS) after major SCI results in grave neurovascular alterations with irreversible damage to the ascending and descending pathways connecting and coordinating the brain and body that can lead to the total paralysis or even to death of patients. In elderly mortality rates after an osteoporotic spinal fracture are comparable to those after femoral neck fractures [2]. There are no fully restorative therapies for SCI as yet.

SCI can have also non-traumatic causes such as degenerative pathologies of the CNS, that can lead to neuronal death and degeneration axons in the gray matter and myelin disruption in the white matter. Degenerative diseases of spine and spinal cord are a major cause of chronic disability in the adult population. The spine and the spinal cord can be subject to infectious diseases or tumors. Awareness of the different manifestations of spinal metastatic disease is essential because the spine is the most common site of osseous metastatic disease (60–70% of patients with systemic cancer will have spinal metastasis).

Reduction of morbidity and mortality associated with spine and spinal cord damage demands of developing the appropriate diagnostic techniques, methods and protocols in order to initiate a timely adequate therapeutic intervention and follow post-intervention process. High resolution non-invasive 3D examination of intact spine and spinal cord morphology including complex structure of the vascular and neuronal network organization is a crucial issue which offers significant advantages to accelerate the process of drug/treatment development.

Variety of well-established techniques such as magnetic resonance (MR), ultrasound (US), positron emission tomography (PET) and X-ray computed tomography (CT) are in daily use for *clinical diagnosis* of spine and spinal cord, but clinical equipment does not yield relevant high resolution and contrast for studying 3D morphology on the cell level.

Trade-off between sub-micrometric resolution required for mapping the architecture of entire neural and vascular networks in spinal cord and large scale of human body pushes scientists to use small animals such as mice for medical research. A wide range of advanced techniques have been developed for studying spinal cord and brain diseases, including tumors, traumatic injury, Parkinson's disease, Huntington's disease, and Alzheimer's disease. However, the intrinsic low contrast among soft tissues of spinal cord for many techniques imposes obstacles in 3D visualization, especially for exploring of unknown morphologies in vivo. The dense material of backbones, which surrounds soft tissue of the spinal cord scatters and attenuates the signal that renders some available techniques inefficient without removal of vertebral bone or its portion.

Generic knowledge of spinal cord microstructure ordinarily relies on the 2D histological slices post mortem. Histology has excellent contrast given by the staining [3, 4], however, the method is destructive, result in shrinkage and deformation of the sample, reconstruction of the 3D geometry of soft and hard tissue from a stack of 2D histological images as well as sample preparation are quite sophisticated and time consuming.

Well-established high-resolution 2D images techniques such as Scanning Electron Microscopy (SEM) complementing with micro-tomography [5] provide impressive 3D image of the vasculature and neuronal system of the cortex and spinal cord. However, this technique is quite complicated and requires an invasive sample preparation, which can alter the 3D morphology of the system.

Magnetic resonance imaging at a microscopic level (micro-MRI) now routinely obtaines resolution of 100 ⌈m³ in living animals, and is achievable down to the scale of 5-10 μm³ in fixed specimens [6, 7]. Micro-MRI is able non-invasively penetrate the bones and permits high resolution 3D longitudinal image of the intact spinal cord [8]. Diffusion MRI and Diffusion Tensor Imaging (DTI) exhibit high contrast of gray and white matter [9, 10] and enable structural tissue determination of normal and injured animal spinal cords with weighted drop injury [11] and myelin-deficient tissues [12]. The main disadvantageous of micro-MRI are insufficient spatial resolution to reveal detailed microstructures of vascular and neuronal network of SC [13], it often requires the contrast medium enhancement, it has long acquisition times.

Positron Emission Tomography (micro-PET) technique is the valuable research tool for 3D viewing of brain cancers and metastases as well as for mapping normal brain function and support drug development in living systems [14-16]. Due to high penetrating radiation micro-PET has practically unlimited depth of imaging within the small animal. On the other hand γ-rays radiation is harmful and may affect tumor size and mimics radiotherapy. PET images have low resolution around 1 mm [14, 15] and for micro-scale structural information a combination of micro-MRI and micro-CT is necessary.

High frequency Micro-ultrasound (micro-US) is a real-time imaging modality capable to study in vivo high speed events such as blood flow and organs function in mice [17]. Micro-US allows noninvasive longitudinal data acquisition in vivo and ex vivo mouse model with a few tenths microns spatial resolution [18-20]. To image micro-vascularization, the resolution can be enhanced to ~5-10 μm with the injection of micro-bubble contrast agents [21, 22]. Although micro-US imaging is harmless, not invasive and low-cost technology, however, it has low acoustic-impedance contrast among soft tissues and a limited imaging depth to a few millimeters for resolution around 20 microns [23, 24].

Fluorescence-based techniques such as confocal laser scanning [25, 26], two-photon and multi-photon excited fluorescence [27, 28] have considerably expanded potential of light microscopy enabling to reveal structural, functional and molecular information on cellular level in vivo. Nanoscopy with fluorescent labeling [29, 30] has allowed reaching resolution up to a few tenths of nanometers. Despite the notable success fluorescence microscopy still have a number of shortcomings related with limited imaging volume, which prevents the detailed 3D visualization of intact spinal cord.

A critical limitation of all optical imaging method is high scattering and attenuation of signal in bony tissue, which make this technique problematic for studies of the spinal cord within intact vertebral column.

X-ray micro-tomography (micro-CT) takes advantage of high penetrating ability of x-ray radiation to maps area of different densities within an object [31]. 3D image is digitally reconstructed from set of radiographs taken from different angles around the sample [32]. Micro-CT is nondestructive relatively fast imaging modality that permits visualization of large tissue volumes at high resolution and contrast between tissues with differential attenuation of X-rays [33]. The technique has been used with great success in small animal ex-vivo and in vivo models [33-36]. Micro-CT provides spatial resolution on micro-scale (5–50μm) (micro-CT) and nano-scale (~100 nanometres) (nano-CT) [35, 37-38] levels and micro-scale features can be readily resolved in opaque samples with dimensions measured from mm to cm [33, 39]. Micro-CT is a powerful tool for imaging of highly absorbing mineralized structures such as bone [34, 40], whereas soft-tissue imaging has low contrast and usually requires the use of X-ray-absorbing contrast agents [41, 42]. Micro-CT combined with contrast agent angiography has been widely applied for investigation of vascularization and angiogenesis of different organs including brain and spinal cord. Synchrotron radiation-based micro-CT (SRmCT) enables improve resolution up to a few microns which allows discriminate the small capillary architecture of cerebral and spinal microcirculation in rodent models [43, 44]. High quality visualization of soft tissue usually requires invasive exogenous contrast [45, 46]. Micro-CT experiment with mouse spinal cord removed from vertebral column and soaked in contrast agent has shown good differentiating between grey and white matter, dorsal horn, ventral horn, dorsal root and ventral root with spatial resolution of a few tenth of microns [47], however the technique is invasive itself and therefore is limited to the postmortem analysis. Less invasive contrast agents enabling a relatively long-lasting contrast in vivo have been proposed for micro-CT analysis of vasculature and organs in a normal and diseased mouse model [48, 49] and for passive tumor targeting of brain cancer [50], but this technique does not allow detailed cell-level visualization of brain and spinal cord morphology [51]. Micro-CT has noticeable advantage in the depth-to-resolution ratio respect to other imaging techniques due to high penetration ability of x-rays. However x-ray radiation is harmful for biological samples and radiation safety is the critical issue for in vivo models, which sets at the moment a limit to obtainable resolution down to a few tenths of a micron for living animals [48, 49]. Other problem of micro-CT is the poor contrast resolution between tissues with similar density, which can be only partially, resolved with invasive contrast agents.

X-ray phase contrast micro-CT (XPmCT) exploits absorption, refraction and scatter contrast generated by phase and intensity variation of X-ray waves transmitted through an object [52]. The main vantage of XPmCT respect to conventional X-ray absorption-contrast techniques is high sensitivity for detecting of small density variations and soft-tissue details without the use of exogenous contrast agents [53]. In the last years, a variety of phase-contrast X-ray imaging techniques such as crystal interferometry [54-55], propagation-based imaging [56-58], analyzer-based imaging [59], edge-illumination [60] and grating-based imaging [61-62] have been developed. High brilliance and low emittance of synchrotron sources have enabled very high resolution 3D visualization of normal and pathological tissues of brain and spinal cord [63-65]. Blood vessels down to the micrometer can be clearly revealed by PCI-CT both with contrast agent added and without any contrast agents [66-72]. Macro- and micro- morphology of intact mice spinal cord with complex organization of vascular and neuronal networks have been shown with sub-micrometer resolution in ex-vivo model [64, 65]. In spite of the obvious success proving the great potential of phase contrast technique, until now visualization of the rodent spinal cord has been limited mainly by imaging of organ extracted from vertebral column. The extremely destructive procedure of bones removal leads to sample deterioration and to loss of considerable part of information about the object. Moreover this procedure is incompatible with investigation in vivo tissue.

We present unprecedented quality sub-micrometric resolution 3D visualization of spinal cord surrounded with intact vertebral column in an ex-vivo mouse model produced with one distance phase contrast synchrotron x-ray micro-tomographic measurement, that allows displaying simultaneously bony tissue, the complete architecture of the VN, neuronal populations, axon bundles up to a single neuron soma, with no contrast agent no sectioning and neither removing the vertebras or other specific sample preparation. The technique takes advantage of high penetration ability of x-rays and benefits as absorption as phase contrasts that permits high resolution, high contrast, 3D volume rendering deep within the body. *Propagation-based phase contrast* is the simplest and extensively used technique because it does not need any optical elements and it works at single acquisition distance. As a consequence, significantly reduction of radiation dose on the sample is possible and therefore this technique is particularly adapted to *in-vivo* biological imaging. Indeed in recent years *in-vivo* phase contrast tomography with few-ten μm spatial resolution at an acceptable dose for small-animal imaging has been performed both for laboratory and synchrotron radiation source [52, 73, 74].
On this regard, our work is an important step toward the in-vivo biological object and clinical implementation.

Advantages of XPmCT based method used in this article concern the combination of wide field of view, sub-micron resolution and high contrast tissue image of unsliced, unstained and hydrated specimens with complex morphology including high and low absorbing components, relatively short image acquisition times. The main disadvantage is radiation damage of biological samples due to ionizing radiation that can pose some limitation of technique for in-vivo model imaging and clinical application.

The additional technical challenges for phase contrast micro-CT of the intact spine and spinal cord are related with problematic imaging conditions, that require the high dynamic range of detector due to extend the X-ray attenuation range between high and low absorbing part of the sample, artifacts and arising noise due to the surrounding bony tissue.

## 2. RESULTS

Standard X-ray Computed Tomography (CT), based on attenuation of hard X-rays in the object, has an excellent absorption contrast for bonny tissue. However the soft tissue of the spinal cord at high energy produces a low absorption contrast in the radiograph and it is usually transparent in a conventional radiography or tomography. Thereby the image of the spinal cord surrounded by the spine bone appears to a challeging task. We combine the advanced X-ray phase contrast tomography in Propagation-Based Imaging (PBI) geometry with developed tools of analysis in order to solve this problem and to achieve detailed high spatial resolution 3D images of the intact spinal cord and the spine.

PCI technique renders phase shifts visible as intensity variations recorded by the detector. Propagation-Based Imaging (PBI) setup consists of an in-line positioned an X-ray source, the sample and an X-ray detector without any other additional optical elements. The selection of the X-ray incident energy and the sample-detector distances has been done taking into account the *transport of intensity equation* in order to optimize the phase contrast produced by the internal structure of the soft spinal cord and the absorption contrast produced by the bone.

Figure 1
X-ray phase contrast tomographic reconstruction of the mice spinal cord (axial view) :
(a,b) the single slice, (c) reconstruction of 50 slices volume of the sample. The image is obtained at incident energy of 60 KeV, pixel size of the detector 3 microns and the sample-detector distance 2.3 ~~1.2~~

m (a,b) and 60 cm (c). The figure shows the detailed structure inside and outside the bone.

Figure 2

X-ray phase contrast tomographic reconstruction of the mice spinal cord:

(a) axial view; (b) longitudinal view. The image is obtained at incident energy 30 KeV, pixel size of the detector 3 microns and the sample-detector distance 2.3 ~~1.2~~ m. The spinal cord is well visible.

Figure 3

X-ray phase contrast tomographic reconstruction of 50 slices volume of the sample and segmentation of the mice spinal cord (axial view). The performed analysis and segmentation produced an impressive treasure of information, both on vascular and neuronal network. The ventral (red arrow) dorsal (blue arrow) median (yellow arrow) dorsal (green), and lateral (orange) spinal arteries are well imaged in white. The details, at the white/grey matter interface, reported in figure 3c and 3d show the micro-vascularization (in black) and the white cells and nerve fibers (in white) respectively. The image is obtained at incident energy 17 KeV, pixel size of the detector 2 microns (a) and 0.64 microns (b-d) and the sample-detector distance 20 cm.

Figure 4

The longitudinal view of the same sample as in Figure 3. The anterior spinal medullary arteries and anterior radicular arteries are well visible (in white) departing from the white matter into the grey matter. The image is obtained at incident energy 17 KeV, pixel size of the detector 0.64 microns and the sample-detector distance 20 cm.

Figure 5

The samples prepared with the additional injection of the microfill (details are reported in the Methods section), a contrast agent able to stain the vascularization. 3D rendering shows the ascending vertebral arteries outside the bone are well imaged and partially filled by the contrast. Also the small vasculature, not filled by the stain, is well visible and rendered in blue in figure 5d, demonstrating that the contrast agent is not necessary to image the vascular network. The image is obtained at incident energy 17 KeV, pixel size of the detector 2 microns and the sample-detector distance 20 cm.

Figure 6

The samples prepared with the additional injection of the iodine. 3D image shows the detailed vascularization inside the white and grey matter of the spinal cord is imaged with high spatial resolution inside the vertebral bone. Artifacts from the high absorbing bone do not affect the richness of the information obtained in the soft matter. The image is obtained at incident energy 60 KeV, pixel size of the detector 1 microns and the sample-detector distance 60 cm.

## 3. MATERIALS AND METHODS

### Theory

Phase Contrast X-ray Imaging (PCI) tomography considers changes in the phase of an X-ray beam that traverses an object. Propagation-Based Imaging (PBI) setup has been used in our experiments to detect phase shifts as intensity variations recorded by the detector remote at some distance from the sample, so the radiation refracted by the sample interferes with the direct beam. The visibility of fringes is proportional to the second derivative of the phase of the wavefront that brings to strong contrast on the structural borders of the sample (so called edge enhancement effect). For complex object containing phase and absorption contrast, the edge enhancement effect is especially strong for an absorption image and the phase information in the image can be lost. There are different approaches that allow retrieve the real and imaginary part of the refractive index separately. However generally the phase retrieval procedure requires several image acquisitions that increase the radiation dose imparted to the specimen.

D. Paganin et al. in Ref. [75] proposed and applied with success a method for the simultaneous phase and amplitude extraction from a single defocused image of a homogeneous object. Apart the fact that the single distance of acquisition significantly reduce radiation load on the sample, an advantage of the methods is its stability with respect to noise, computational speed and simplicity of implementation. In Ref. [76, 77] authors extend the work of Paganin *et al*. [75] to enable analytic propagation-based phase-retrieval tomography to be performed on a multi-material object. The algorithm use a single PBI image per projection and separately reconstructs each interface between any given pair of distinct materials.

Our paper deals with the X-ray phase contrast micro-tomography of complex biological object (mouse spine) contained both high contrast absorption part (dorsal vertebras) and low contrast transparent part

(spinal cord). To reconstruct the internal structure of the object a set of phase contrast projections from a single distance PBI is taken at different illumination angles of the object. The phase retrieval of projected image was made on the basis of Paganin method (PM) and PM extended for multimaterial object. The set of retrieved images consequently was used for 3D computer reconstruction of internal structure of the object.

Image contrast of weakly absorbing spinal cord tissue can be significantly enhanced while the artifact given by highly absorbing bones and imperfect experimental conditions can be reduced with the next image processing procedures:

      Sinogram preprocessing.

      Regularization in reconstruction method.

      Post-processing of the reconstructed data.

Generally, the sinogram preprocessing is undesirable operation since it can lead to essential unpredictable deformation of geometry of the reconstructed object. However, we have had to use sinogram filtering in order to reduce ring- and metal- artifacts. Ring artifact is caused by a miscalibtarted or defective detector elements as well as by the uncontrolled change in the intensity or/and the profile of the x-ray beam during the long-time measurements. Ring artifacts were suppressed via normalization procedure so that the background corresponding to maximum intensity of transmitted beam for all frames remains constant. Background level was determined from the area in which direct beam does not overlaps the object. If no such areas, for example in the ROI tomography, normalization has been done so that the Radon integral has been constant for all projections. These procedures allows significantly reduce artifact without deformation of the geometry of the reconstructed object.

The metal artifacts look like bright rays diverging from the strong absorbing object. They are caused ether by starvation of photons passed through the strong absorbing areas or/and by a low dynamic range of the detector. Often they make interpretation of reconstructed image complicated or even impossible. In order to reduce artifacts we applied the adaptive filtration of a sinogram decreasing the impact of strong absorbing peaks. Since in these areas the signal-to-noise ratio is low, the above filtration preserves information we are interested in about weakly absorbing structure. Since there were an adequate number of projections with low signal-to-noise ratio, we took advantage of the FBP method without resorting to algebraic reconstruction techniques. Finely we have applied a threshold filtration of the bone tissue in the reconstructed image which allows further increase the contrast of weakly absorbing spinal cord and other soft tissue relative to the entire object.

## 4. CONCLUSIONS

The future of the investigation of alterations in CNS, due to injuries or pathological conditions, is the high spatial resolution in-vivo studies. In particular, the monitoring of advanced treatments and therapies would require high spatial resolution investigation tools. In this work we present an original data analysis procedure to get 3D images of mice spine, preserving the richness of micro-details of the spinal cord inhabiting inside. Our results are the first step forward to the difficult way toward the sub-micro in-vivo investigation.

**FIGURE 1**

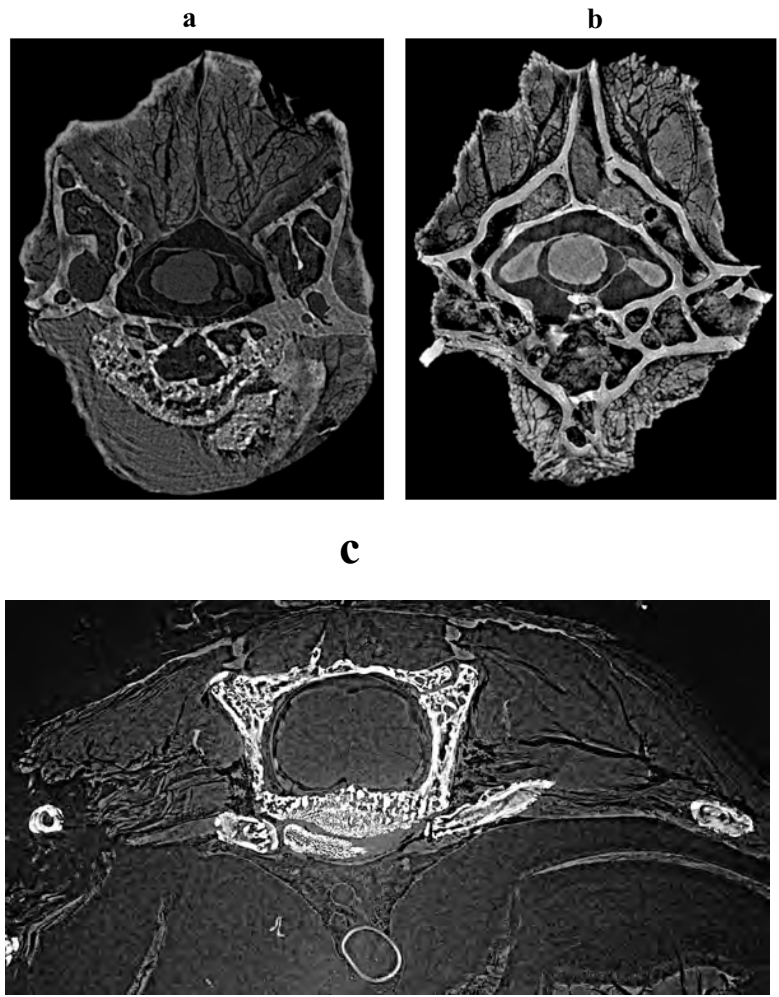

**FIGURE 2**

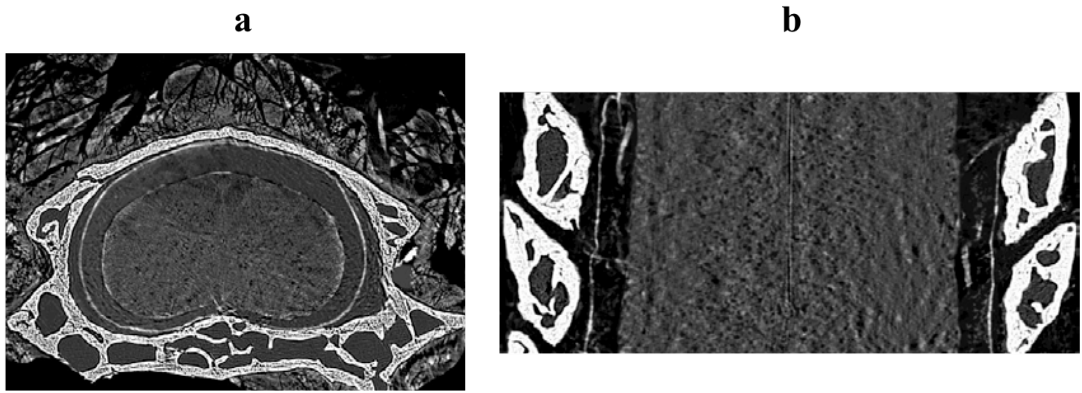

**FIGURE 3**

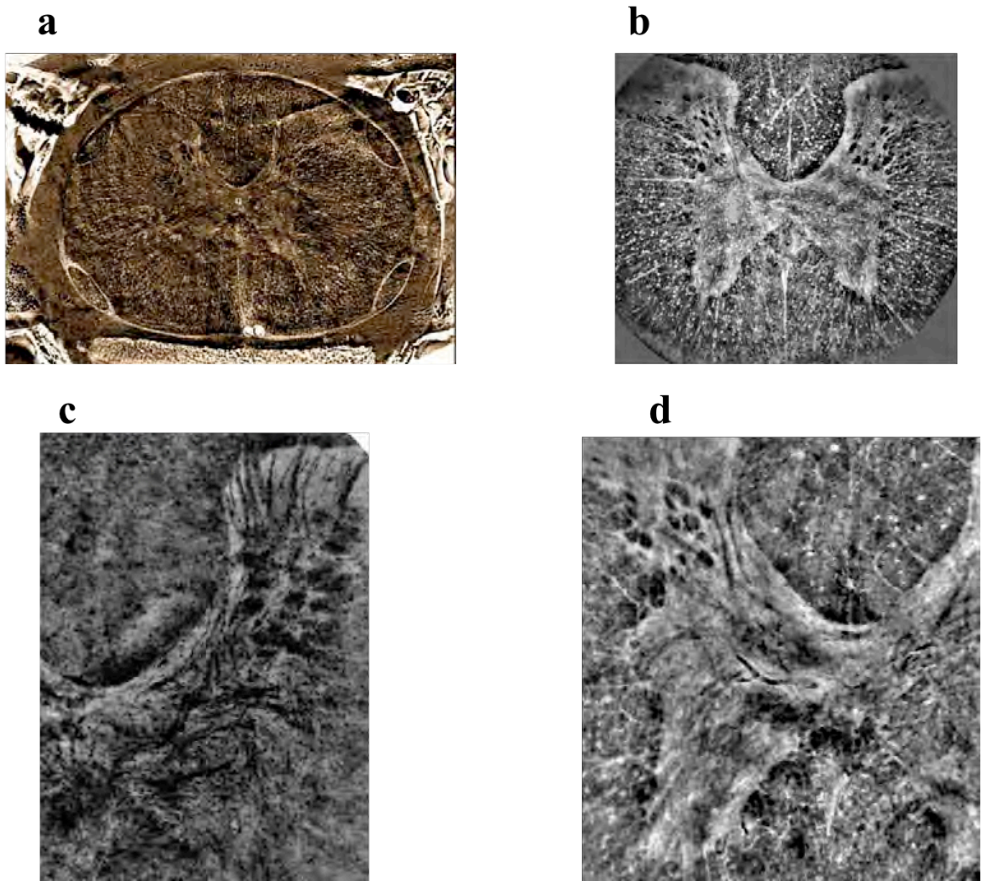

**FIGURE 4**

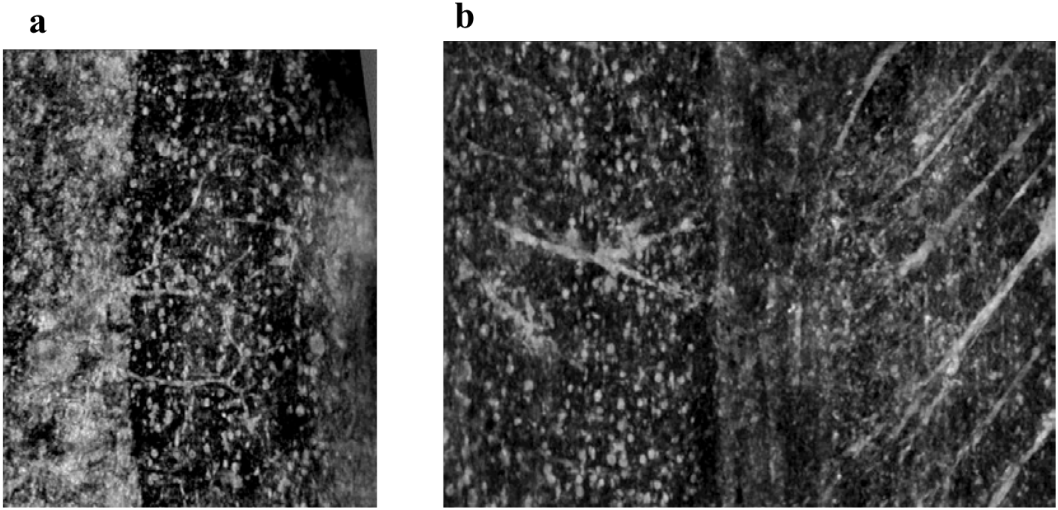

**FIGURE 5**

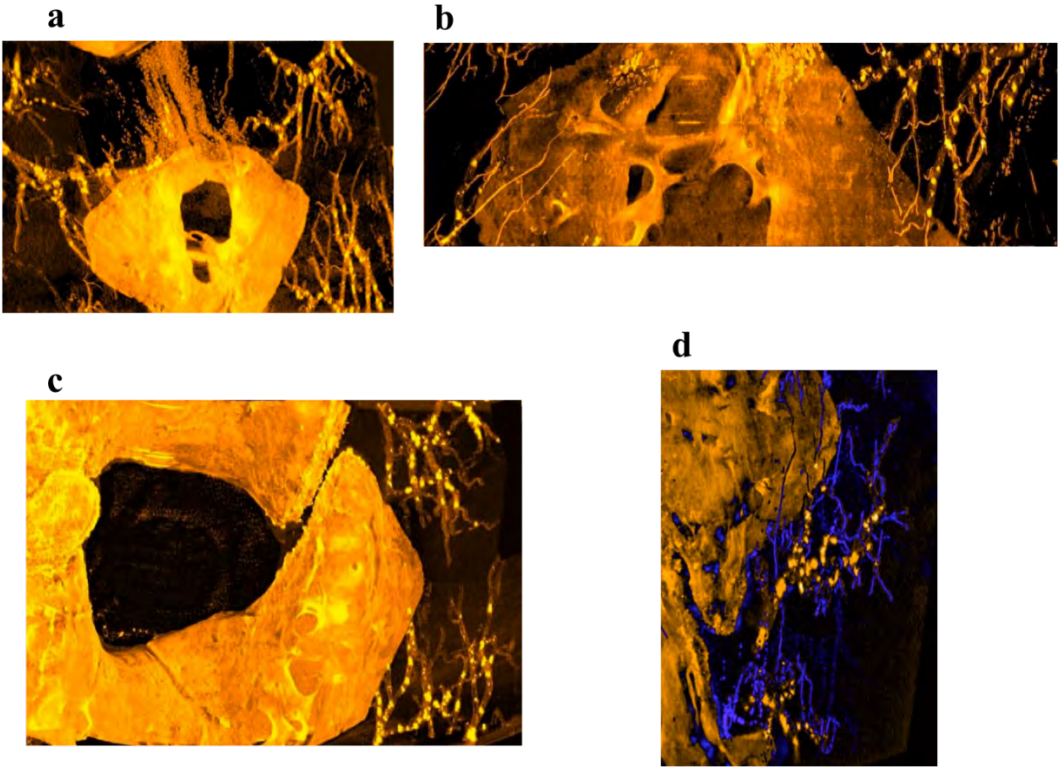

**FIGIRE 6**

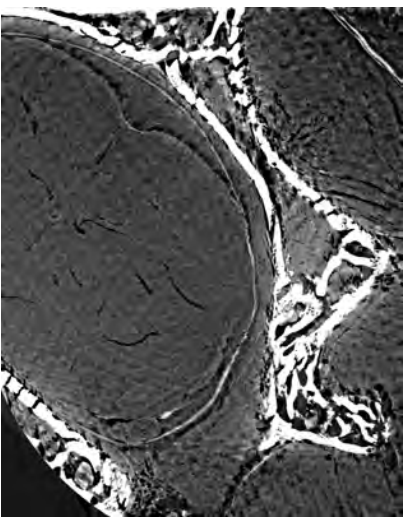